\documentstyle[preprint,tighten,eqsecnum,aps,epsf]{revtex}

\draft

\def\d{{\rm d}}
\def\w{\omega}

\newcommand{\teta}{\rlap{\lower2ex\hbox{$\,\tilde{}$}}\eta{}}

\newcommand{\T}{\textstyle}

\tighten
\begin{document}
\preprint{\vbox{
\rightline{SU-GP-99/6-1}
\rightline{NSF-ITP-99-061}
\rightline{gr-qc/9906078}}}
\title{On a spacetime duality in $2+1$ gravity}

\author{Alejandro Corichi\thanks{Electronic Address:
corichi@nuclecu.unam.mx}${}^{1,2}$ and
Andr\'es Gomberoff\thanks{Electronic Address:
agombero@phy.syr.edu}${}^{2,3,4}$}
\address{1. Instituto de Ciencias Nucleares, Universidad Nacional 
Aut\'onoma de M\'exico\\ A. Postal 70-543, M\'exico D.F. 04510,
 M\'exico}
\address{2. Institute for Theoretical Physics, University of 
California\\ 
Santa Barbara, CA 93106, USA}
\address{3. Physics Department, Syracuse University\\
 Syracuse, NY 13244, USA}
\address{4. Centro de Estudios Cient\'{\i}ficos de Santiago\\
 Casilla 16433, Santiago 9, Chile}

\date{June,1999}
\maketitle

\begin{abstract}

We consider $2+1$ dimensional gravity with a cosmological constant,
and explore a duality that exists between space-times that have the
De Sitter group $SO(3,1)$ as its local isometry group.  In
particular, the Lorentzian theory with a positive cosmological
constant is dual to the Euclidean theory with a negative
cosmological constant.  We use this duality to construct a mapping
between apparently unrelated space-times.  More precisely, we
exhibit a relation between the Euclidean BTZ family and some
$T^2$-cosmological solutions, and between De-Sitter point particle
space-times and the analytic continuations of Anti-De Sitter point
particles.  We discuss some possible applications for BH and $AdS$
thermodynamics.

\end{abstract}

\section{Introduction}
\label{intro}

In order to capture some qualitative features of $3+1$ dimensional
gravity, it is always useful to consider simpler toy models that are
free of many of the technical difficulties of the full theory.  This
is particularly true of $2+1$ gravity.  Before 1984, roughly
speaking, the theory was considered to be ``too trivial'' to deserve
any attention.  It was with the work of Deser, Jackiw and t`Hooft
\cite{deser84},\cite{deser2}, who considered point particle
solutions, and later with the papers by Achucarro and Townsend
\cite{achuca} and Witten \cite{witten1}, that an avalanche of papers
on $2+1$ gravity followed.  A very wide spectrum of issues have been
addressed in this years, from the ``problem of time'' to the ``loop
representation'', and more recently, black hole solutions
\cite{BTZ1}.  For recent reviews see \cite{carlip1} and particularly
\cite{carlip2}.  For a review of the $2+1$ black hole see
\cite{carlip3}.  $(2+1)$--dimensional gravity has been successful in
providing simplified models in the study of black hole physics.  In
particular, the case when a negative cosmological constant is added
(anti--de Sitter space-times) has been extensively studied in the
last years.  On the one hand it has been shown \cite{BTZ2} that
making certain identifications in anti--de Sitter (AdS) geometry,
one obtains a Black Hole solution (BTZ black hole).  On the other
hand, in string theories it has been found that there is a whole
family of supersymmetric black brane solutions whose ``near
horizon'' geometry is the product of spheres and the three
dimensional AdS spacetime \cite{branes}.  The study of the quantum
properties of the BTZ black hole has been studied using many
approaches, all of them giving a quantum mechanical derivation of
the black hole entropy \cite{carlip?},\cite{max},\cite{strominger}.

The case of a positive cosmological constant has also received
recent attention.  Maldacena and Strominger \cite{malda:strom} have
given a CFT derivation of the entropy associated to the cosmological
horizon, following earlier results by Carlip \cite{carlip4}.  More
recently, similar derivations in the Euclidean continuation have
given an alternative explanation of the Hawking-Gibbons entropy of
De Sitter space-time \cite{max:3,lin:wu}.

In this note we consider three dimensional gravity with a
cosmological constant.  The purpose of the paper is to explore a
little-known duality that exists between 
Lorentzian $2+1$ gravity with a {\it
positive} cosmological constant and Euclidean gravity with a {\it
negative} constant \cite{aa:ac}.  
As first shown in \cite{achuca}, $2+1$ gravity
can be formulated as a theory of connections.  There are two ways
of writing the 3-dimensional action that involve connections.  The
Einstein-Hilbert-Palatini action has as independent fields a
``spin'' connection $\omega$ and a (non-degenerate) co-triad field
$e$.  The ``pure-connection'' formulation considers a Chern-Simons
action where $\omega$ and $e$ are combined into a single connection
$A$ living in a larger gauge group.  This later description
simplifies the structure of the theory and it is possible to
completely characterize the state space.  When one works in the so
called ``frozen formalism'', i.e., in the reduced phase space of the
theory, one is dealing with the ``true degrees of freedom'' of the
theory.  That is, roughly speaking, an equivalence class of
classical solutions where one does not distinguish between
configurations that are related by a ``gauge transformation''.
However, in the cases of interest for this note ($(-++)$ signature
 with $\Lambda>0$ and $(+++)$ with $\Lambda < 0$), both
the space of histories and the reduced phase
spaces have been shown to {\it coincide} \cite{aa:ac}.
  Thus, we have a mathematical identical
description of the two systems.  This is quite surprising and
disturbing at first.  But, as it is the case in many physical
situations, the ``physics'' is in the interpretation of the
formalism.  As we shall see, there is a precise way to reconstruct
two space-times, each one with a different signature, for each point
of the (reduced) phase space.  This construct also provides an
explicit mapping between this space-times, or using a
``Wheelerism'', a {\it Wick rotation without Wick rotation}.

Let us illustrate this duality using some heuristic arguments.
Take the Euclidean BTZ solution with positive mass $M$ and
no angular momentum\footnote{We are using units such that $8G=1$. 
In three 
spacetime dimensions this means that mass is dimensionless and 
$\hbar$ has units of length.},

\begin{equation}
ds^2 = \frac{(r^2-M l^2)}{l^2} d\tau^2 + 
\frac{l^2}{r^2 - M l^2} dr^2 + r^2 d\phi^2 \ ,
\label{BTZ}
\end{equation}
 
and make the following identifications:
\begin{equation}
l \longrightarrow il\qquad ; \qquad \tau
\longrightarrow i\tau \, .
\label{analitic}
\end{equation}
We obtain,
\begin{equation}
\d s^2 = - \frac{l^2}{T^2 + M l^2} \d T^2  + \frac{(T^2+M 
l^2)}{l^2} \d R^2
+ T^2 \d\phi^2 \ .
\label{btzcont}
\end{equation} 

Here we have changed the names of the variables $r\rightarrow  T$ 
and $\tau 
\rightarrow R$.

This solution looks like a cosmological ``big--bang'' solution, 
with a 
singularity at $T=0$.  In the next sections we will see that this 
is
indeed the case, and that this space-time is in a precise sense, 
dual
to the BTZ solution.

Now take (\ref{BTZ}) with $M = -\alpha^2 \in [-1,0]$. These are 
``particle'' 
solutions, with no horizon and naked, conical singularities at 
$r=0$. The   
angle deficit of this cones is $\Omega=2\pi(1-\alpha)$.
For $\Omega=0$ we get the 
singular free global AdS spacetime.  Proceeding with 
(\ref{analitic}) in this solution we get

\begin{equation}
\d s^2 = -\frac{(l^2\alpha^2 -r^2)}{l^2} \d\tau^2 +
\frac{l^2}{l^2 \alpha^2 - r^2} \d r^2 + r^2 
\d\phi^2 \ .
\label{dspart}
\end{equation}
This represents a spacetime with a cosmological horizon at
$r=r_c=l \alpha$ and 
a conical singularity at $r=0$ with deficit angle $\Omega$.
Near  the horizon, defining $\rho^2 =  l^2 -r^2/\alpha^2$, we 
find,
\begin{equation}
ds^2 \approx -\frac{\alpha^2\rho^2}{l^2} d\tau^2 + 
d\rho^2 + \alpha^2 l^2 d\phi^2 \ .
\end{equation}
Note that in order that the Euclidean continuation of this 
solution is 
singularity free in the $(\rho,\tau)$ plane, we must identify the 
time $\tau$ 
with period $\beta = 2\pi l/\alpha$. This  period can be 
associated to the  inverse temperature of the solution, 
which in this case is \cite{max:3},
\begin{equation}
T =\frac{\alpha\hbar}{2\pi l} = \frac{r_c\hbar}{2\pi l^2} \ .
\end{equation}

It is worth noting that we could formally associate a temperature 
to the 
dual of the particle case, which had no horizon and thus
no natural notion of temperature. 
Reciprocally, the originally hot BTZ black hole is continued to a
cosmological solution with no natural temperature, for which one 
could
again assign some formal temperature. 

The structure of the paper is as follows. In Sec.~\ref{sec:2}
 we recall the gauge theory formulation of $2+1$ gravity
with a cosmological constant, and 
the equivalence between the choices of signature and cosmological
constant $\Lambda$ that we are
interested in. In Sec.~\ref{sec:3} we explore in detail this duality, and
apply it to the interesting case of the BTZ black hole `family'.
In particular, we give a rigorous justification for the 
`dualities'
presented above. We end with a discussion in Sec.~\ref{sec:4}. In the 
Appendix,
we consider the case of non-zero angular momentum.

\section{preliminaries}
\label{sec:2}

In this section we recall some basic features of $2+1$ gravity in 
the
presence of a cosmological constant. The Einstein-Hilbert action
 for 3-dimensional gravity,
\begin{equation}
 S_{\rm EH}[g]:=\frac{1}{2\pi}
\int_M \d^3\!x\,\sqrt{|g|}(R-2\Lambda)\, ,
\end{equation}
can be rewritten in terms of a $1$--form connection $\omega$ and a 
$1$-form triad field $e$ as follows \cite{ash-rom,romano},
\begin{equation}
S_{\rm P}[\omega,e]:=
\frac{1}{2\pi} \int_M \epsilon_{IJK}
\left(F(\omega)^{IJ}\wedge e^K -\frac{\Lambda}{3} e^I\wedge e^J 
\wedge e^K 
\right)\, ,
\label{actionX}
\end{equation}
where the capital indices denote internal vectors, that is, 
vectors in
a fiducial 3 dimensional vector space $W$.
 The field $e^I_{\mu}$ is the
soldering form that maps internal vectors to tangent vectors to 
the
3 manifold $M$: $v^I:=e^I_\mu v^\mu$. In the internal vector space 
$W$ there is
a fixed fiducial metric $g_{IJ}$ whose signature coincides with 
the one of the space-time metric. Therefore, in the case of 
Euclidean 
gravity
$g_{IJ}=\delta_{IJ}$ and for Lorentzian signature it is the 3 
dimensional
Minkowski metric $\eta_{IJ}$ with signature $(-,+,+)$.
 There is an Lorentz invariant (or rotation invariant in the 
Euclidean case) 
volume element $\epsilon_{IJK}$ in $W$ defined from $g_{IJ}$. The 
$\epsilon$ 
with ``upstairs'' indices is
defined from $\epsilon_{IJK}$ by ``raising the indices'' with 
$g^{IJ}$. 
We can recover the space-time metric $g_{\mu\nu}$ from the soldering 
form $e_\mu^I$,
\begin{equation}
g_{\mu\nu}:= e_\mu^I e_\nu^J g_{IJ}\, .   
\end{equation}
Finally, $F^{IJ}_{\mu\nu}$ is the curvature 
$2$--form  of the connection 
$\omega^{IJ}_{\mu}$,{\em 
i.e.}, 
$F^{IJ}=\d\omega^{IJ}+\omega^I_{\ K}\wedge\omega^{KJ}$.

The variation of (\ref{actionX}) with respect to $\omega_{IJ}$ yields,
\begin{equation}
\d e^{I} + \omega^{I}_{J} \wedge e^J=0\, ,\label{ele1}
\end{equation}
and the equation from $e$ is,
\begin{equation}
F^{IJ} - \Lambda e^{I}\wedge e^{J} =0.\label{ele2}
\end{equation}
The first equation states that the torsion of $\omega$ vanishes,
and implies  that the covariant derivative defined by 
${\omega_{\mu I}}^J$ coincides with the one compatible with  the 
co-triad, that is,
${\omega_{\mu I}}^J={\Gamma_{\mu I}}^J$. We can therefore replace 
$F$ by 
$R$ in
 (\ref{ele2}) and we arrive at Einstein equations:
\begin{equation}
G^{\mu\nu}+\Lambda g^{\mu\nu}=0\, .\label{einst}
\end{equation}
Recall that this equation implies that the space-time has constant 
scalar curvature proportional to $\Lambda$:
\begin{equation}
g_{\mu\nu}(R^{\mu\nu}-\frac{1}{2}Rg^{\mu\nu}+\Lambda 
g^{\mu\nu})=-\frac{1}{2}R+3\Lambda=0\, .
\end{equation}
Therefore, $R=6\Lambda$. Note that this result is independent of 
the
signature of the space-time and therefore, of the gauge group we 
are considering.

At this point it is convenient to use a fact from 3 dimensions: 
the
dimensions of the ``Lorentz group'' and the manifold coincide. 
That is, the  Lie 
algebra $so(2,1) (so(3))$ and the internal space $W$ can be 
identified.
Furthermore, the Killing-Cartan metric on the Lie algebra $k_{IJ}$
 is proportional to the internal metric $g_{IJ}$.  Note that the 
connection is a 
``Lie-Algebra valued'' $1$ --form, and  its components are labeled 
by two 
vectorial indices (the natural labels in the vectorial or defining  
representation). However, in $3$ dimensions, the adjoint 
representation and the 
defining representation of $so(2,1) (so(3))$ coincide. It is 
simpler to work 
with $1$--index labels (the natural ones in the adjoint 
representation). We therefore define
\begin{equation}
{\omega_{I}}^J =: \, {\epsilon^J}_{IK}\omega^{K}
\label{isom}
\end{equation}
 It is in this step that the Euclidean and
Lorentzian theories have different expressions. This can be 
understood from
the fact that we are using the internal metric $g$ to raise the 
indices of the $\epsilon_{IJK}$ in equation (\ref{isom}):
${\epsilon^J}_{IK}:=g^{JM}\epsilon_{MIK}$.
We can now rewrite the generalized covariant derivative in terms 
of $\omega_\mu^I$ as,
\begin{equation}
{\cal D}_\mu v^I=\partial_\mu v^I+[\omega_\mu,v]^I,
\end{equation}
where $[\omega_\mu,v]^I={\epsilon^I}_{JK}\,\omega_\mu^J\,v^K$. 
{}From (\ref{isom}) it also follows that the generalized curvature
tensor two--form  $F^I$ is,
\begin{equation}
F^I=\d\omega^I+\frac{1}{2}{\epsilon^{I}}_{JK}\omega^J \wedge 
\omega^K.
\end{equation}

The Palatini action can be rewritten as,
\begin{eqnarray}
S_L[e,\omega]&=&
\frac{1}{2\pi}\int_M \left(2F(\omega)_I\wedge e^I 
-\epsilon_{IJK}\frac{\Lambda_L}{3} e^I\wedge e^J 
\wedge e^K 
\right)
\qquad \;\;{\rm Lorentzian,}\label{actionL}\\
S_E[e,\omega]&=&-\frac{1}{2\pi}\int_M \left(2F(\omega)_I\wedge 
e^I+\epsilon_{IJK}\frac{\Lambda_E}{3} e^I\wedge e^J 
\wedge e^K 
\right)
\qquad{\rm Euclidean.}
\label{actionY}
\end{eqnarray}

The actions (\ref{actionL}) and (\ref{actionY}), in spite of the
fact that are defined for different Lie groups and have
different signs, are in a precise sense, equivalent.
This can be seen at different levels (for details see \cite{aa:ac}).
The easiest way of showing their equivalence is to generalize
the observation of Witten \cite{witten1}
that one can define a new connection
$A^a$ from the old connection $\w$ and the co-triad $e$, living
on a larger Lie algebra. 
This generalization, originally due to Romano
\cite{romano}, can be applied to any action that has the structure
of actions (\ref{actionL}) and (\ref{actionY}).
It is defined, in particular, for  any semi-simple
Lie group $G$ and depends on a free parameter
$\kappa$.\footnote{The type of  action is of a
Palatini-$BF$  form:
$S_{BF}=\int{\rm tr}( B\wedge(2F+\frac{\kappa}{3}B\wedge B))$.}.
The result is that any such action is equal to a Chern-Simons
action based on the $\kappa$-extension, $\kappa G$, of the
original group $G$. For the cases of interest, that is, for the
groups $SO(3)$ and $SO(2,1)$, their $\kappa$-extension coincide,
being the group $SO(3,1)$, when $\kappa=-1$. The gravity actions
can be put in the $BF$ form if we construct the $B$ field from
the drei-bein $e$ as follows: 
$B^I_\mu:=\sqrt{|\Lambda|}\,e^I_\mu$. Then, the
gravity actions (\ref{actionL}),(\ref{actionY}) give rise to
the {\it same} Chern-Simons action
when $\Lambda_L > 0$ and $\Lambda_E=-\Lambda_L$.
For details of this equivalence, manifested also at the phase
space level see \cite{aa:ac}. The other cases give rise to
the groups $\kappa G=SO(2,2)$ when $G=SO(2,1)$ and $\Lambda_L<0$,
and $\kappa G=SO(4)$ when $G=SO(3)$ and $\Lambda_E>0$.

In this note we shall restrict our attention to 
the cases in which the $\kappa$-extension corresponds to
the {\it De-Sitter} group $SO(3,1)$\footnote{Our use of the term 
De Sitter
is sometimes different from what has been used in the literature
(see for instance \cite{max:3}). For us, a {\it De-Sitter 
spacetime},
Euclidean or Lorentzian is one for which the (local) isometry 
group is the De-Sitter group $SO(3,1)$.}.  
Solutions to the equation of the motion coming from the
Chern-Simons action are flat connections,
and two solutions related by a gauge  transformation
are regarded as physically indistinguishable. Thus, the physical
phase space $\hat{\Gamma}$ is the moduli space of flat $SO(3,1)$
connections or, in other words, the homomorphisms from $\pi_1(M)$
to $SO(3,1)$ (modulo the adjoint action of the group). 

There are several remarks. First, a point in $\hat{\Gamma}$
represents both (an equivalence class of)  Lorentzian space-times
 (3-geometries)
with a {\it positive} cosmological constant and Euclidean
space-times with a {\it negative} cosmological constant! Thus,
there exists, at the fundamental level, a mathematical equivalence 
between
the two theories. It is precisely this duality that we shall
explore in this note. Second, it is no coincidence that the group
$SO(3,1)$ appears as the Chern-Simons gauge  group, since it is 
also
the global symmetry group of the De-Sitter (dS) space
in three dimensions. In fact, one knows that a space-time of the
desired topology can be found by appropriate identifications 
and/or
quotients of the dS space by a subgroup of the $SO(3,1)$ 
group  \cite{witten1}. 

The relation between the  elements of the group $G$,
associated to every non-contractible loop, and the quotient 
construction
is what allows us to reconstruct a spacetime given a set of group
elements. Furthermore, one is able to reconstruct {\it two} 
space-times:
one Lorentzian ($\Lambda > 0$) and one Euclidean ($\Lambda < 0$). 
For,
the global De-Sitter solutions can be viewed as embedded in four
dimensional Minkowski space-time; the Lorentzian De-Sitter 
(${\rm dS}_{\rm L}$) is the hyperboloid defined by 
$x_0^2+x_1^2+x_2^2-x_3^2=l^2$, where $l^2=1/|\Lambda|$, and the
Euclidean De-Sitter (${\rm dS}_{\rm E}$) can be taken as one
of  the hyperboloids defined by $x_0^2+x_1^2+x_2^2-x_3^2=-l^2$.
It is easy to
see that the scalar curvature of these space-times is positive
for ${\rm dS}_{\rm L}$ and negative for ${\rm dS}_{\rm E}$, so
they are indeed solutions to the Einstein equations (\ref{einst}).

It is important to note that this ``holonomy" duality is, in a 
sense,
complementary to the well known Wick rotation (analytic 
continuation).
In this later case, one in changing the signature of the embedding 
space
so the  symmetry group of the global,  maximally extended, 
geometry is
not preserved, whereas the sign of the cosmological constant, and 
therefore
the scalar curvature, is preserved. For instance, the Wick 
rotation of
Anti De Sitter (AdS), a Lorentzian space-time with $\Lambda < 0$ 
and
group $G=SO(2,2)$ is
Euclidean De-Sitter ${\rm dS}_{\rm E}$ for which $\Lambda < 0$ and
$G=SO(3,1)$. 
 
\section{Dual Space-times: some examples}
\label{sec:3}

In this section, we construct explicitly the duality mentioned
 in previous sections for a particular family of (very well known)
 solutions: the BTZ family. In the Euclidean side this includes 
the
 BTZ black hole (of positive mass), going through the zero mass 
solution
 and the so called ``mass gap" finishing with the global Euclidean 
De-Sitter.
 We shall study their Lorentzian duals and comment on their 
thermodynamic
 properties.

\subsection{De Sitter space-times}
\label{sec:3.a}

\subsubsection{Lorentzian De Sitter}

The Lorentzian De Sitter Spacetime  ${\rm dS}_{\rm L}$
can be defined as the surface
\begin{equation}
x_0^2 + x_1^2 + x_2^2 - x_3^2 = l^2 \ \ ,
\end{equation} 
embedded in the four dimensional Minkowski spacetime with line 
element,
\begin{equation}
\d s^2= \d x_0^2 + \d x_1^2 +\d x_2^2 -\d x_3^2 \ \ .
\end{equation}

The topology of the space-time is $S^2\times R$.
This spacetime is maximally symmetric, i.e., it has six Killing 
vectors (three rotations and three boosts),
which form a $SO(3,1)$ isometry group.

A standard choice of coordinates $(t,\theta,\phi)$
that cover the entire space-time are
\cite{haw:ellis},
\begin{eqnarray}
x_0 = l \cosh t \sin\theta\sin \phi \quad &;&\quad
x_1 = l \cosh t \sin\theta \cos \phi \\ \nonumber
x_2 = l \cosh t \cos\theta \quad &;& \quad
x_3 = l \sinh t \label{global}
\end{eqnarray}
The induced metric then takes the form,
\begin{equation}
\d s^2=l^2\left[-\d t^2+ \cosh^2t(\d 
\theta^2+\sin^2\theta\d\phi^2)
\right]\label{4.4} 
\end{equation}
where $t\in (-\infty,\infty)$, $\phi\in[0,2\pi)$, $ 
\theta\in[0,\pi)$.
As can be easily seen, the spacetime is a two sphere that 
contracts
to its minimum area $4\pi l^2$ at $t=0$ and then expands again.

In spite of the convenience of working on a globally defined 
coordinate
patch, we shall introduce a popular choice of coordinates in which
the metric looks ``static". First let us introduce some other 
embedding
coordinates that will cover  the region $x_0^2+x_1^2 \le l^2$ and
$x_2\ge 0$,

\begin{eqnarray}
x_0 = l \sin \chi \sin \phi \quad &;&\quad
x_1 = l \sin \chi \cos \phi \\ \nonumber
x_2 = l \cos \chi  \cosh \tau \quad &;& \quad
x_3 = l \cos \chi  \sinh \tau \label{4.5} 
\end{eqnarray}
Here $\phi \in [0,2\pi)$, $\chi \in [0,\pi/2]$ and $\tau \in 
(-\infty,\infty)$. 
The 
metric induced in this region is,

\begin{equation}
\d s_{I}^2 = -l^2  \cos^2\chi \ \d\tau^2 + l^2\d\chi^2 +
l^2 \sin^2\chi \ \d\phi^2 \ . 
\label{part1}
\end{equation}

Defining the coordinates $r=l\sin \chi$ and $t=l\tau$ we get,
\begin{equation}
\d s_{I}^2 = -\left(1-\frac{r^2}{l^2}\right)\d t^2+
\left(1-\frac{r^2}{l^2}\right)^{-1}\d r^2+r^2 \d\phi^2
\end{equation}
where $t\in(-\infty,\infty)$, $\phi\in[0,2\pi)$, $r\in [0,l]$.
In this ``Schwarzschild" coordinates, $r=0$, the origin, is the
location of an observer and $r=l$ is its cosmological (event)
horizon \cite{haw:gibb}. Using standard arguments in which
one analytically continues the solution to a Euclidean (in this 
case, locally isometric to
 the three sphere $S^3$) spacetime, and requires
regularity in the $(r,\tau)$ plane,
one can assign a temperature to
the horizon to be \cite{haw:gibb2},
\begin{equation}
T_{h}=\frac{\hbar}{2\pi l}
\end{equation}
This temperature is just ($1/2\pi$ times) the surface gravity of 
the
(properly normalized) killing vector generating the horizon
\cite{haw:gibb}.
In the global coordinate system (\ref{global}) the observer is
located at the ``north pole" of the two-sphere ($\theta=0$). The
antipodal point would be represented, in a Penrose-Carter diagram, 
as living in another ``asymptotic region" \cite{haw:gibb}, and
is not covered by the coordinate patch (\ref{4.5}).

\subsubsection{Euclidean De Sitter}

The Euclidean De Sitter space-time ${\rm dS}_{\rm E}$ 
can be defined as the surface
\begin{equation}
x_0^2 + x_1^2 + x_2^2 - x_3^2 = -l^2 \ \ ,\label{4.13}
\end{equation} 
embedded in the four dimensional Minkowski spacetime with line 
element,
\begin{equation}
\d s^2= \d x_0^2 + \d x_1^2 +\d x_2^2 -\d x_3^2 \ \ .
\end{equation}

This space-time is disconnected, each connected component being a
hyperboloid. It is standard to choose one of them, say, the upper
one as the space-time.
The topology of the space-time is $R^3$.
This spacetime is maximally symmetric, i.e., it has six Killing 
vectors (three rotations and three boosts),
which form a $SO(3,1)$ isometry group.

A standard choice of coordinates $(\tau,\theta,\phi)$
that cover all the spacetime ($x_3\geq 0$) are, 
\cite{haw:ellis},
\begin{eqnarray}
x_0 = l \sinh \tau \sin\theta\sin \phi \quad &;&\quad
x_1 = l \sinh \tau \sin\theta \cos \phi \\ \nonumber
x_2 = l \sinh \tau \cos\theta \quad &;& \quad
x_3 = l \cosh \tau \label{global2}
\end{eqnarray}
The induced metric then takes the form,
\begin{equation}
\d s^2=l^2\left[\d \tau^2+ \sinh^2\tau(\d 
\theta^2+\sin^2\theta\d\phi^2)
\right]
\end{equation}
where $\tau\in [0,\infty)$, $\phi\in[0,2\pi)$,
$\theta\in[0,\pi)$.
The point $\tau=0$ is a regular point and is just a singular point 
of the coordinate system.

This space-time can also be put in a static form.
For that we need to
introduce some new embedding
(covering the same region of spacetime) as
follows,
\begin{eqnarray}
x_0 = l \sinh \rho \sin \phi \quad &;&\quad
x_1 = l \sinh \rho \cos \phi \\ \nonumber
x_2 = l \cosh \rho \sinh t^{\prime} \quad &;& \quad
x_3 = l \cosh \rho \cosh t^{\prime}\label{4.17}
\end{eqnarray}
The induced metric then takes the form,
\begin{equation}
\d s^2=l^2\left[\cosh^2\rho \d {t^{\prime}}^2+ 
\d \rho^2+\sinh^2\rho\d\phi^2
\right]\label{4.18}
\end{equation}
where $\rho\in [0,\infty)$, $\phi\in[0,2\pi)$, 
$t^{\prime}\in (-\infty,\infty)$.

Defining the coordinates $r=l\sinh \rho$ and $t=lt^{\prime}$ we 
get,
\begin{equation}
\d s_{I}^2 = \left(1+\frac{r^2}{l^2}\right)\d t^2+
\left(1+\frac{r^2}{l^2}\right)^{-1}\d r^2+r^2 \d\phi^2
\end{equation}
where $t\in(-\infty,\infty)$, $\phi\in[0,2\pi)$, $r\in 
[0,\infty)$.
This is precisely the ``Wick transform" of (a patch of)
the global Anti--de Sitter 
(AdS) spacetime through the transformation $t\mapsto it$.

\subsection{The Boost Sector: BTZ Black Hole vs. Cosmological
$T^2$}

\subsubsection{Euclidean Signature}

In this part we construct, in an explicit fashion, a 
Euclidean BTZ black hole of mass $M$. We do this in order to
illustrate the procedure that will lead to the (one parameter
family of) dual Lorentzian cosmological space-times. For simplicity
we shall consider the non-rotating black hole. The rotating case
can be found in the Appendix. 
The starting point is Euclidean De Sitter ${\rm dS}_{\rm E}$ 
(\ref{4.13}). 
Next, we consider the Killing vector field,
\begin{equation}
\xi=x_2\frac{\partial}{\partial x_3}+ x_3\frac{\partial}{\partial 
x_2}
 \ \ ,
\end{equation}
which generates boosts in the plane $(x_2,x_3)$ and it is tangent 
to the
De Sitter space. Note that $|\xi|^2=x_3^2-x_2^2=l^2+x_0^2+x_1^2$. 
  Let us now define the embedding (\ref{4.17}), changing 
the names of the coordinates as follows,
\begin{eqnarray}
x_0 = l \sinh \rho \sin \chi \quad &;&\quad
x_1 = l \sinh \rho \cos \chi \nonumber\\
x_2 = l \cosh \rho \sinh \varphi \quad &;& \quad
x_3 = l \cosh \rho \cosh \varphi \label{btz:emb}
\end{eqnarray}
we have the line element,
\begin{equation}
\d s^2=l^2\left[\sinh^2\rho \,\d \chi^2+ 
\d \rho^2+\cosh^2\rho\,\d\varphi^2
\right]
\label{met}
\end{equation}
where $\rho\in [0,\infty)$, $\chi\in[0,2\pi)$, 
$\varphi\in (-\infty,\infty)$. Note that 
$\xi=\frac{\partial}{\partial \varphi}$, and $\varphi$ is an 
affine
parameter of the finite ``boost" along the integral curves of
$\xi$.
The metric  (\ref{met}) induces naturally the topology $R^3$ over 
the
region 
covered by the embedding (\ref{btz:emb}). Note however, that from a
canonical 
perspective, where we require the space-time to be
 of the form $M=\Sigma \times R$, we are forced to  take
out the point $\rho=0$
from the space-time, and as a consequence of that, the topology
will change to $S^1\times R^2$. We shall construct the solution and 
see
that in the BTZ case, one can add the point $\rho=0$
to the space `generated' via the Hamiltonian $2+1$ evolution 
and still have a regular space-time with a different topology.

The BTZ solution is constructed by periodically identifying
$\varphi$ with $\varphi + \varphi_0$, where
$\varphi_0=\frac{2\pi r_{+}}{l}$. Here $r_+$ is
simply a parameter labeling the space-time. What one does is to
compactify along  $\varphi$. If we consider the
point $\rho=0$ to be part of space-time
we have a $S^1\times R^2$ topology, and if
we remove it, we  end up with a
space-time of topology $T^2\times R$ (needed from the canonical
perspective). 
Note that even when we pick this later choice,
we are respecting the $2\pi$ periodicity
of the $\chi$ coordinate. This will translate into a trivial 
holonomy
along its (artificially constructed, topological nontrivial) orbits. 
Let us see that we indeed get the BTZ solution. Writing,
$\chi= (r_+/l^2) \tau$ and $\varphi=(r_+/l)\phi$, we have
\begin{equation}
\d s^2=\frac{r_+^2}{l^2}\sinh^2\rho \,\d {\tau}^2+ 
l^2\d \rho^2+r_+^2\cosh^2\rho\,\d\phi^2
\label{btzrho}
\end{equation}
where $\rho\in [0,\infty)$, $\tau\in[0,\frac{2\pi l^2}{r_+})$, 
$\phi\in [0,2\pi)$.
Now, changing coordinates $r=r_+\cosh\rho$ we get,
\begin{equation}
\d s^2=\left(\frac{r^2}{l^2}-\frac{r_+^2}{l^2}\right) \d \tau^2+
\left(\frac{r^2}{l^2}-\frac{r_+^2}{l^2}\right)^{-1} \d r^2+r^2\d 
\phi^2
\label{BTZe}
\end{equation}

We see that the horizon is at $r=r_+$ ($\rho=0$) and the $\tau$
coordinate, the parameter along the Killing field 
$\frac{\partial}{\partial\tau}$ is naturally periodic with period
$(2\pi\l^2)/r_+$. From here we see that the temperature that we
should assign is,
\begin{equation}
T_{\rm bh}=\frac{\hbar r_+}{2\pi l^2}
\end{equation}
Finally, the mass of the space-time is 
$M:=\frac{r_+^2}{l^2}$, where the value  $M=0$ is assigned, as 
usual, 
to the 
solution with $r_+=0$. 

The following questions come to mind:
what is the relation between this construction and the reduced
phase space in terms of $\pi_1(M)\rightarrow SO(3,1)$?
In what sense can we say that the holonomies along the
orbits of $\xi$ are related to the parameter $r_+/l$?
 The relation between the appearance of this non-trivial 
holonomies  and 
the construction described above deserves more attention. 
The procedure involves the construction of a 3-dimensional
space-time $M$, out of a point of the reduced phase space.
This procedure naturally implies some sort of ``gauge fixing"
in order to select a representative form the equivalence class
of flat $SO(3,1)$ connections, that will have a corresponding
 space-time locally isometric to $M^4$. To see that the space-time
constructed 
via identifications has associated to it a flat
$SO(3,1)$ connection whose holonomies parameterize the
reduced phase space we need to use the structure available in
$M^4$. The $SO(3,1)$ 3-d connection is constructed out of the 
structure defined on
the embedding space-time, namely, the flat $SO(3,1)$
connection compatible with the Minkowski metric\footnote{This
$SO(3,1)$ connection parallel transports vectors with
internal indices. In Minkowski space-time there is a canonical,
globally defined
soldering form that translates internal vectors into
space-time vectors. We shall use this structure to refer
to internal vectors as simply `vectors'}
and  the
action $\phi$ of the group $SO(3,1)$ acting on $M^4$
(i.e., $\phi:SO(3,1)\rightarrow {\rm Diff}(M^4)$, $g\mapsto
\phi(g):M^4\rightarrow M^4$). 
The hyperboloids
corresponding to De-Sitter space-times are invariant subspaces of 
the
$SO(3,1)$ action $\phi$, and the group acts transitively. That is, 
given
any two points $p$ and $q$ on $M$, we can always find an
element $g\in SO(3,1)$
such that $(\phi(g))(p)=q$. Let us now define a canonical
(auxiliary) 3-d flat connection
 on $M$ as follows: Given an open curve
$\gamma$  on $M$, with $p$ as its starting point and
$q$ as its end point, and an initial vector $\vec{v}(p)$, we
 parallel transport the vector using the flat $SO(3,1)$
connection of the embedding space-time.

Thus, the connection so defined is a true $SO(3,1)$ connection on 
$M$.
Furthermore, given that the action $\phi$ of $SO(3,1)$ on $M^4$
 is linear, the derivative $\phi_{*}$ coincides with it, and is also
given by an $SO(3,1)$ action on tangent vectors.
Thus, if we parallel transport
a 3-dimensional vector (using the natural connection)
along the orbits of the
KVF $\xi$ from the origin $\varphi=0$ to say, the point $q$ 
defined by $\varphi=\varphi_0$, the
original vector and the parallel transported one will be the ``same" 
(for instance in a Cartesian coordinate system).
How can we say that the parallel
transported vector will be different from the original one when we 
are using the flat constructed above?
Since we are identifying $q$ with the origin, we are at 
that step closing the curve $\gamma$ into a  loop, and we 
have to ``bring back" the vector to the starting point. That is,
we ``push" forward the vector from $q$ to $p$ using the
mapping $(\phi_*(g))^{-1}$.  Thus,
the element of the group $SO(3,1)$ that provides the parallel
transport (in this case, a boost), becomes the holonomy along the
non-contractible loop $\gamma$.
Therefore the ``parallel transported'' vector has as its
holonomy precisely the group element in $SO(3,1)$ corresponding
to a boost along $\xi$ with parameter $\varphi_0$. This is the 
relation between holonomies and space-time identifications.
To see more about the relation between geometry
and holonomies (known as geometric structures) see \cite{carliph}. 

Now, the topology of the spacetime is $T^2\times R$ and therefore
we have two non-contractible generators of $\pi_1(M)={\bf Z}_2$.
The gauge invariant information, i.e., the coordinates of 
$\hat{\Gamma}$ are four parameters $(a_1,s_1,a_2,s_2)$ 
\cite{ezawa,aa:ac},
where $a_i$ are rotation parameters corresponding to a holonomy
along the $[\gamma_i]$ generator of $\pi_1(M)$, and $s_i$ correspond
to boost parameters along the dual plane in $M^4$. 
Concretely,  representatives of the $[\gamma_1]$ equivalence class
are given by the closed orbits of the $\partial/\partial\tau$ vector 
field and representatives of  $[\gamma_2]$
are given by the orbits of $\partial/\partial\phi$. 
In this language,
a (Euclidean) BTZ black hole of mass $M$ and $J=0$ has coordinates 
in $\hat{\Gamma}$ equal to $(2\pi,0,0,(2\pi r_+)/l)$.

\subsubsection{Lorentzian Signature}

In order to construct the holonomy-duals to the BTZ black hole, we 
will  proceed 
analogously. First, we take the {\it same} killing vector
\begin{equation}
\xi=x_2\frac{\partial}{\partial x_3}+ x_3\frac{\partial}{\partial 
x_2}
 \ \ ,
\end{equation} 
and identify points in ${\rm dS}_{\rm L}$ 
space along this killing vector. This 
procedure will induce closed time-like curves in the region where
$|\xi|^2 < 0$, therefore we will take out this region from the 
spacetime
and take the surface  $|\xi|=0$ as a ``chronological 
singularity''.  

The region of interest its covered by the following patch of 
coordinates 

\begin{eqnarray}
x_0 = l \cosh t \sin \chi \quad &;&   \quad
x_1 = l \cosh t \cos \chi \\ \nonumber
x_2 = l \sinh t \sinh \varphi \quad &;& \quad
x_3 = l \sinh t \cosh \varphi 
\end{eqnarray}
with  $-\infty<t,\varphi<\infty$ and $0 \le \chi < 2\pi$. In these 
coordinates 
the 
metric takes the form,

\begin{equation}
\d s^2 = -l^2 \d t^2 + l^2\cosh^2 t \ \d\chi^2 + l^2 \sinh^2 t \
\d\varphi^2 \ .
\label{bang}
\end{equation}

We can now safely identify $\varphi$ with $\varphi + 2\pi b$, 
which makes the surfaces 
of constant time compact $T^2$ manifolds. Notice that in order
to relate it to the BTZ black hole we have the correspondence 
$b \leftrightarrow r_+/l$. 
The chronological singularity $|\xi|^2=0$ is 
now located at $t=0$, which is a naked ``initial'' singularity. 
For big values 
of $t$, the spatial slices have the topology of a 2--torus whose 
radius are 
expanding exponentially in time. The parameter $b$, labeling 
the
particular space-time,  gives 
the ratio between the two radii. 
For small $t$ one of the radius goes to $l$ 
while the other shrinks to zero (see Figure 1).

In order to put this solution in a familiar form set
$\chi=(b/l) {\chi}^{\prime}$, $\phi=(1/b)\varphi$
 and $T=lb \sinh t$. Thus, we
have,
\begin{equation}
\d s^2=-\left(b^2+\frac{T^2}{l^2}\right)^{-1}\d T^2+
\left(b^2+\frac{T^2}{l^2}\right) \d {{\chi}^{\prime}}^2
+T^2\d\phi^2\label{ezawa}
\end{equation}
where $T\geq 0$, ${\chi}^{\prime}\in [0,(2\pi l)/b)$, and
$\phi\in[0,2\pi)$.

These solutions are not new. They are a subclass of the general
solution found by Ezawa \cite{ezawa} (see also
\cite{carlip1} and \cite{mess}). It is important to note
that in terms of the reduced phase space language, these solutions
have coordinates $(2\pi,0,0,2\pi b)$, and that the 
holonomy-duality is relating  space-times (\ref{BTZe}) and
(\ref{ezawa}) when we set $b=r_+/l$.

\begin{center}
\leavevmode
\epsfxsize=12cm
\epsfbox{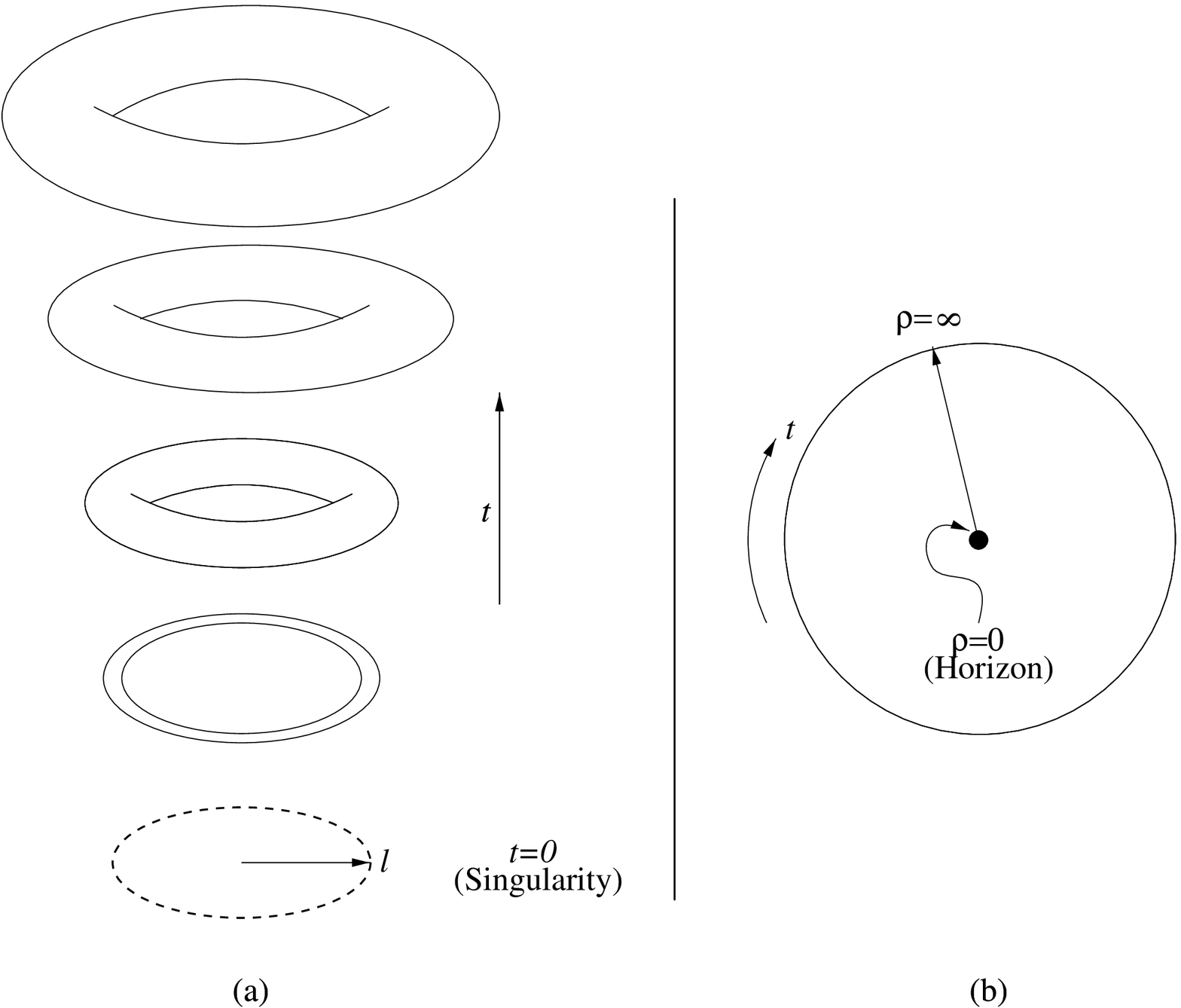}
\end{center}

FIGURE 1: The ``Big Bang'' spacetime (a) vs. The BTZ spacetime 
(b). In (b) we represent the plane $(t,\rho)$ in the coordinates 
given by the metric (\ref{btzrho}). Each point of this plane 
has a circle on it, which completes the spacetime (with its 
$R^2\times S^1$ topology. $r_+$ represents the radius of the 
circle at the origin. In (a), the parameter $b$ defines the geometry 
of
the $t=$ constant torus.
The duality identifies $b l$ with $r_+$.

\subsection{The Rotation Sector: Particle Solutions}

In this part we shall consider another class of solutions.
Just as the $T^2$-BTZ-cosmological solutions were found by
means of an identification along a boost, the solutions in this 
section
will be constructed using a rotation.

\subsubsection{Lorentzian Signature}

Point particle solutions for De-Sitter spacetime were studied by 
Deser and
Jackiw \cite{deser2}. They are a generalization of the point 
particle
solutions for flat 3-dimensional gravity studied in 
\cite{deser84}.
Recall from Sec.~\ref{sec:3.a} that the global De Sitter has a
$S^2\times R$ topology.
Thus, as argued by Deser and Jackiw, one cannot construct
a one-particle space-time by removing a ``wedge" from the $S^2$ 
spatial
slice since there will be, at least, two conical singularities. In 
our
coordinate system $(t,\theta,\phi)$ (\ref{4.4}) the two particles 
will
be located at antipodal points on the sphere ($\theta=0$ and 
$\theta=\pi$).
However, the Schwarzschild patch (\ref{4.5}) only sees one of the
observers/particles and the other one is hidden behind the 
cosmological
horizon. The `remove a wedge' construction of Deser and Jackiw can 
be
interpreted in our language as, again, an identification along the
orbits of a Killing field. The main difference with the BTZ 
construction
is that now we take a rotation KVF 
$\zeta=\frac{\partial}{\partial\phi}$
instead of a boost. When one uses
a rotation field to construct a new spacetime,
 one needs due care. For, one cannot simply
identify, say, a point with angle coordinate $\phi$ with
the point $\phi+\phi_0$, for $\phi_0 < 2\pi$. This naive quotient 
procedure
 leads (for almost all values of $\phi_0$) to a space that is not 
even
a manifold. Instead, the construction involves the arbitrary 
choice
of a value of, say $\phi=0$, such that the points $\phi=0$ and
 $\phi=\phi_0$ are to be
identified and the wedge between $\phi_0$ and $2\pi$ is removed 
from the
manifold. Even though the construction procedure is different from 
the
one used in the BTZ case, the picture of having an element of 
$SO(3,1)$,
the holonomy, for each non-contractible loop remains unchanged.
This is so since by cutting the wedge, one has two 
curvature-singular
points at the poles of the sphere that have to be removed from the
manifold. Thus, the resulting topology of $M$ is $S^1\times R^2$ 
and
we have one non-trivial loop around the $S^1$. 
As an end result, we have that the holonomy along this loop, the 
(new)
orbit of the vector $\partial_\phi$, is a rotation (in SO(3,1)) 
by $\phi_0$.  

In this case we take the region of ${\rm dS}_{\rm L}$ space 
covered
 by the  patch of coordinates  described below.

\begin{eqnarray}
x_0 = l \sin \chi \sin \phi \quad &:& \quad
x_1 = l \sin \chi \cos \phi\nonumber \\
x_2 = l \cos \chi  \cosh \tau \quad &;& \quad
x_3 = l \cos \chi  \sinh \tau  \label{3.26}
\end{eqnarray}

Here $\phi \in [0,2\pi)$, $\chi \in [0,\pi/2]$ and $\tau \in 
(-\infty,\infty)$. 
The 
metric induced in this region is,

\begin{equation}
\d s_{I}^2 = -l^2  \cos^2\chi \ \d\tau^2 + l^2\d\chi^2 +
l^2 \sin^2\chi \ \d\phi^2 \ . 
\label{part2}
\end{equation}

Cutting the wedge means restricting the range of $\phi$ to 
$[0, 2\pi \alpha)$
and identifying $\phi=0$ with $\phi=2\pi\alpha$, where $\alpha\in 
(0,1]$. 
Thus, the space-time
has a deficit angle of $\Omega=2\pi(1-\alpha)$. When $\alpha=1$, 
there
is no deficit and we recover global De Sitter.
If we now change coordinates to $\varphi=\phi/\alpha$, 
$\tau=(\alpha/l) t$ and
$r=l\alpha\sin\chi$ we get,
\begin{equation}
\d s_{I}^2 = -\left(\alpha^2-\frac{r^2}{l^2}\right)\d t^2+
\left(\alpha^2-\frac{r^2}{l^2}\right)^{-1}\d r^2+r^2 \d\varphi^2
\label{3.32}
\end{equation}
where $\varphi\in[0,2\pi)$, $r\in [0,\alpha l]$ and 
$t\in(-\infty,\infty)$.
The cosmological horizon is now located at $r_c=\alpha l$ and, 
following the
argument given in the Introduction, the temperature
of the horizon is $T_{h}=\alpha\hbar/(2\pi l)$.
The deficit angle is manifested by the fact that the `area' of the
horizon is now $A_{h}=2\pi \alpha l$.

\subsubsection{Euclidean Signature}

Let us now consider the Euclidean case. The starting point is 
Euclidean
De Sitter spacetime (\ref{4.13}) with topology $R^3$. Now it is 
possible
to have a single point particle located at the `origin' $\rho=0$. 
if we
remove this world-line, we are left with a space-time with 
topology
$S^1\times R^2$, with a single non-contractible loop.
We shall use the
same rotational KVZ $\zeta$ as in the case of Lorentzian signature
to identify along its orbits. The resulting space-time, with 
deficit
angle $\Omega=2\pi(1-\alpha)$ is given by,
\begin{equation}
\d s^2 = \left(\alpha^2+\frac{r^2}{l^2}\right)\d \tau^2+
\left(\alpha^2+\frac{r^2}{l^2}\right)^{-1}\d r^2+r^2 \d\phi^2
\label{3.33}
\end{equation}
which corresponds to the Wick transform of the `point particle'
Anti-De Sitter spacetime of mass $M=-\alpha^2$.
 Just as we did for the $T^2$-cosmological
space-times, we could, formally, assign a temperature to this 
point
particle solution to be equal to $T_p=\alpha\hbar/(2\pi l)$.

\subsection{The Vacuum}

This spacetime is the limit $r_+\mapsto 0$ in both Lorentzian and
Euclidean Signatures. 

For the Euclidean signature,
we can simply take the limit of (\ref{BTZe}) when $r_+\mapsto 0$,
\begin{equation}
\d s^2=\frac{r^2}{l^2} \d \tau^2+
\frac{l^2}{r^2} \d r^2+r^2\d \phi^2
\label{vacce}
\end{equation}
now, $\tau$ is no longer a periodic coordinate (recall that its 
range,
for the case $r_+>0$ was $\tau\in[0,\frac{2\pi l^2}{r_+})$),
 so we have a space-time
with topology $S^1\times R$ and zero temperature. Note that we 
would
have arrived at the same space-time in the limit $\alpha \mapsto 
0$
in (\ref{3.32}). In this case, $\alpha$ had the interpretation of 
the range of $\phi$ after  one has removed the wedge
 from the spacetime, so the
limit $\alpha \mapsto 0$ might be interpreted as the limit in 
which
the conical singularity 'opens up' to form a cylinder, thus 
recovering the $S^1\times R$ topology.
This space-time is referred as the vacuum since it is the Wick 
rotation
of the BTZ black hole of zero mass and this is considered to be 
the
zero energy solution \cite{BTZ2}.

The $r_+\mapsto 0$ limit of out Lorentzian solutions has also some
similar features. That is, starting from the $T^2$-cosmological 
solutions (\ref{ezawa}), and taking the limit $b\mapsto 0$ we get,
\begin{equation}
\d s^2=-\frac{l^2}{T^2} \d T^2+
\frac{T^2}{l^2} \d \chi^2+T^2\d \phi^2
\label{vaccl}
\end{equation}
where $\chi$ is not periodic and it takes values in 
$(-\infty,\infty)$,
$T\geq 0$ and $\phi\in[0,2\pi)$. It is again a space-time with
topology $S^1\times R^2$. Starting
from the point-particle family (\ref{3.32}), it corresponds to 
taking
the limit $\alpha\mapsto 0$, where the cosmological
horizon shrinks to zero, and it can  be interpreted
as the 2-sphere opening up to a cylinder. Again. the temperature
we would assign to this vacuum state would be zero.

\section{Discussion and Outlook}
\label{sec:4}

The main results of this note can be summarized in Fig.2. Here, we 
have
the four possibilities of signatures and signs of the cosmological
constant. In the left column we have Lorentzian space-times and in 
the
right their corresponding analytic continuations (Wick 
transforms),
as shown by the `W' arrows (for Wick). We
have listed the space-times considered in this note. The `Holonomy'
transform (H-duality) corresponds to the diagonal `H'
arrows going from the  lower-left 
corner to the upper right corner and relating the  space-times 
studied in this note.

\vskip 1cm
\begin{center}
\epsfxsize=12cm
\leavevmode
\epsfbox{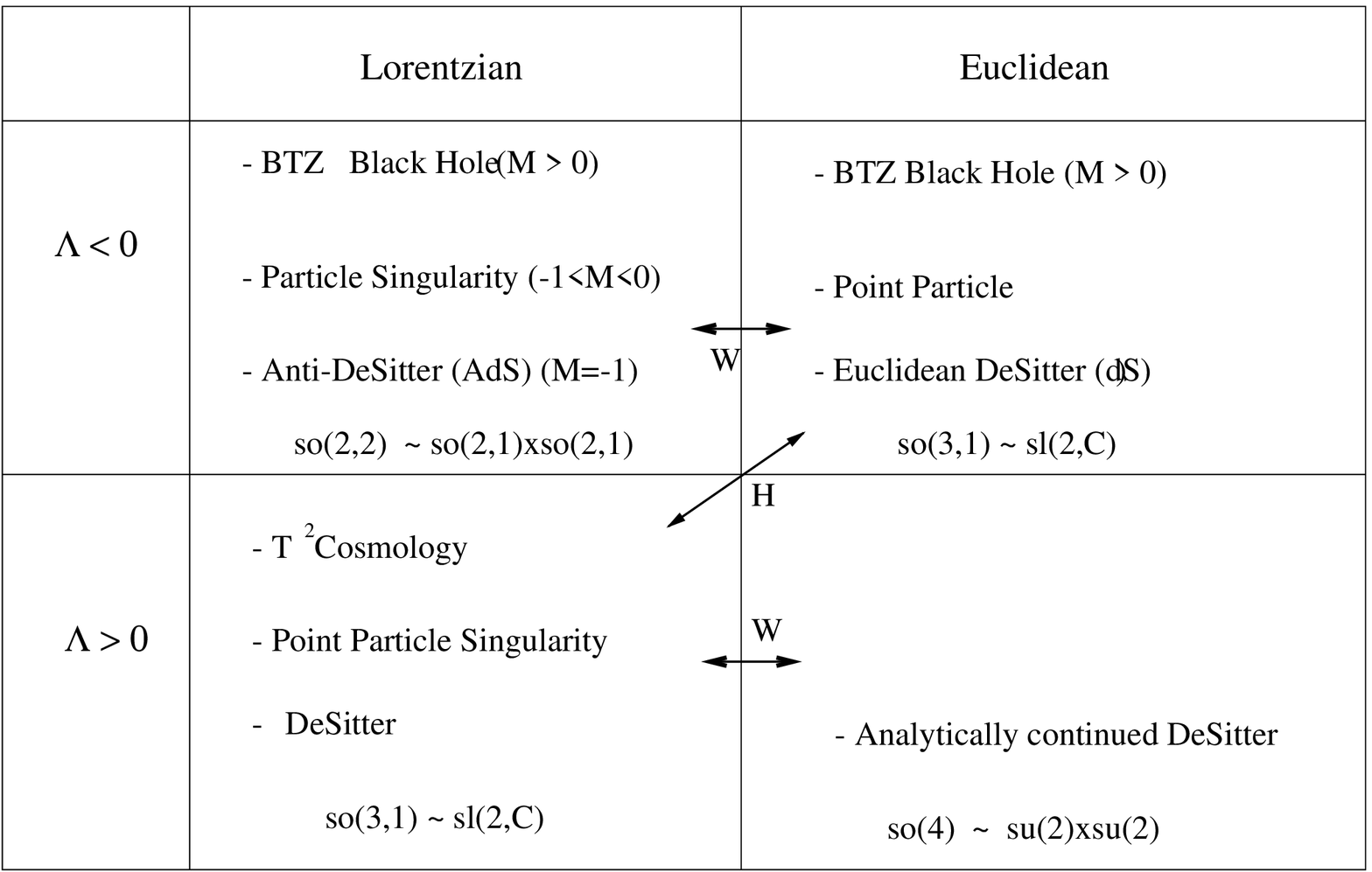}
\end{center}

There are several remarks that need to be made.
\begin{enumerate}

\item {\it Relation to Wick Transform}. As  already discussed, the
standard
Wick transform and the Holonomy duality are in a sense, 
complementary.
First, the Wick-duality for space-times is restricted to complex
space-times that admit both an Euclidean and a Lorentzian real
section. Normally this is possible only in the presence of a
time-like KVF. Furthermore, the mapping in not strictly between
globally defined space-times, since, for instance, it is only the
exterior region of the Lorentzian BTZ black hole that gets ``mapped"
to the Euclidean sector. On the other hand, the Holonomy transform,
being a duality at the phase space level, relates globally
defined space-times. Furthermore, the existence of a KVF is not
essential for the definition of the duality. The Wick rotation
has been extensively used in gravitational physics. Not only is
it used for QFT calculations, but also in connection with
gravitational thermodynamics. Standard arguments of Hawking and 
Gibbons
\cite{haw:gibb2} assign a temperature to the Lorentzian space-time
by looking at the period of the Wick-rotated time in the dual
space-time. This physical equivalence between space-times is used
regularly in the study of thermodynamics of black holes. One may
wonder if the Holonomy dual space-times can be regarded as having
more than just the phase-space equivalence. Is it justified
to assign thermodynamical properties to the dual space-time to
define, say, a temperature and entropy for space-times that do not 
even
have a horizon? Can we perform a calculation in one domain and
conclude something about the other domain just as we do with
the Wick transform?

Affirmative answers to this intriguing questions would indeed
be of some interest since we could, for instance, pass from
the physics in $AdS$ to $dS_L$ composing a $W$ and a $H$-transform.
What would  the relevance be for an
hypothetical $dS/CFT$ conjecture remains an open question.

\item{\it Thermodynamics}. 
Let us elaborate on the possible assignment of a temperature to the
space-times studied in previous section. Let us first recall the
standard argument to define a physical temperature for a Euclidean
space-time. First, we start we a Lorentzian stationary
black-hole space-time, with an $N-1$-dimensional (null)  horizon 
(with
$N$ being the dimension of space-time). In
the Euclidean sector, the horizon is now  $N-2$ dimensional and 
`time'
is a periodic coordinate. One asks for regularity of the space-time 
at
the horizon in order to fix the periodicity of the Euclidean time.
Even if we start with a Euclidean solution with non-trivial 
topology,
the fact that it is regular is not enough to define a temperature.
We need, as in the Schwarzschild black hole case,
the Lorentzian time that {\it measures
proper time at infinity} in order to have the `right' temperature.
The same is true for the  de-Sitter
space-time where one needs to choose the time-parameter as the
proper time of the observer whose cosmological horizon we want to
consider. The periodicity of the Wick-transformed time is then 
related
to the temperature. 

Let us now consider the space-times of interest for this note. The
(non-rotating) Euclidean BTZ solutions have a
naturally defined temperature $T_{\rm bh}=\hbar r_+/(2\pi l^2)$.
On the other hand, the $H$-dual space-time
is a spatially closed space-time, with no
time-like KVF and no horizon. One may wonder what, if any, would be
the meaning of formally
defining thermodynamic parameters and, in particular,
a {\it constant} temperature to the whole space-time.
We shall return to this question shortly.

The Lorentzian De-Sitter space-time together with the point particle
solutions (\ref{3.32}) have horizons with a temperature (assigned 
via
the Wick rotation) given by $T_{\rm h}=\hbar\alpha/(2\pi l)$. The
dual space-times (\ref{3.33}) do not have a periodic `time' 
coordinate
$\tau$ and therefore, no natural temperature. However, if we declare
the temperature of this space-times to be $T_{\rm 
p}:=\hbar\alpha/(2\pi
l)$,
(i.e., the temperature of its $H$-dual) then the natural
period for $\tau$ is $2\pi l/\alpha$. Now, the space-times 
(\ref{3.33})
are the Wick transform of the family of point particle $AdS$ 
space-time
(with $M\in [-1,0)$). Therefore, using the standard Wick transform
argument,
we can assign a temperature to the family of point-particle-$AdS$
space-times to be,
\begin{equation}
T_{\alpha}:=\frac{\hbar\alpha}{2\pi l}
\end{equation}
Note that with this definition, the $AdS$ space-time (corresponding
to $M=-1$ and $\alpha=1$) is a `thermal state' with temperature
$T_{AdS}=\hbar/(2\pi l)$,
recovering the suggestion of Strominger and others
\cite{strominger,malda:stro2}. 


In recent years there have been several attempts to calculate the
entropy of the BTZ black hole.  One of those calculations,
originally due to Carlip \cite{carlip4} (see also \cite{BBO}),
starts by considering the $SL(2,C)$ Chern-Simons phase space for a
$T^2\times R$ space-time.  Then, one imposes boundary conditions,
motivated by the Lorentz black hole horizon, and computes the
Bekenstein-Hawking entropy using the WZW theory induced on the
horizon.  One intriguing fact about this calculation is that, in the
Euclidean domain, one does not have the geometrical interpretation
of the Lorentzian horizon, and without the knowledge of its origin,
we would only be faced with some boundary conditions on phase space
inducing a WZW theory on the boundary.

Since the $H$-duality is precisely based on the fact that the two
gravity theories share the same phase space, then we could perfectly
well interpret the result of the calculation as giving us the
entropy of the Lorentzian cosmological space-time.  The
correspondence is given as follows.  The horizon $\rho=0$ of the
Euclidean black hole corresponds in the Lorentzian side to the
initial $T=0$ singularity (this is a striking property, because
$T=0$ is a {\em chronological} singularity and the horizon $\rho=0$,
is just a coordinate singularity).  This singularity is a mild one,
since it is manifested by the fact that the spatial slice is a
degenerate torus.  If would be of interest to geometrically
interpret the boundary conditions in the Lorentzian theory and to
identify the calculated entropy with some geometrical invariant, the
analogous of the horizon area in the BH case.

Another interesting calculation refers to the entropy
associated to the cosmological horizon in De-Sitter. The first
approach, very much in the style of Carlip is given in
\cite{malda:strom}. Maldacena and Strominger
 again impose some boundary conditions
in the $SL(2,C)$ Chern-Simons phase space and
compute the Bekenstein-Hawking 
 entropy using the CFT on the boundary of the region,
in this case, the horizon. The computed entropy is, as expected,
one fourth of the cosmological horizon radius, $S=2\pi l/(4G\hbar)$.
The other calculation \cite{max:3} computes the Gibbons-Hawking
entropy by considering the partition function in the Wick-rotated
theory (corresponding to a Chern-Simons theory on $so(3)\times 
so(3)$).
The Maldacena-Strominger results can be re-interpreted
in the same way that Carlip's results were (see also
\cite{malda:stro2}).
 That is, one can
argue that the calculation very well applies to the $H$-dual theory,
namely, to a Euclidean De-Sitter space-time with topology $S^1\times 
R$
(\ref{3.33}), and assigns to it an entropy $S:=2\pi 
l\alpha/(4G\hbar)$.
Now, recall that these space-times are the Wick-duals of the point
particle $AdS$ space-times, and therefore one is naturally lead to
assign to these space-times an entropy as well.  For the
global $AdS$ the associated entropy would be
$S_{AdS}:=2\pi l/(4G\hbar)$.
Just as for the cosmological space-times, we would like to interpret
this entropy in geometrical  terms.

\item{\it Path Integrals, Wick Rotations, etc..}

The fact that three dimensional gravity has a dual description, both
as a theory of triads and connections $(\w,e)$ and as a theory of
space-time metrics, poses a puzzle.  On the one hand, there is the
not so unpopular wisdom that there is a deep relation between the
Lorentzian and Euclidean path integral formulation of gravity.
Independently of the viewpoint one adopts, be it that the Euclidean
path integral is useful for calculating Lorentzian transition
amplitudes, or that the Euclidean regime is fundamental, there is
some belief that Euclidean methods are relevant for quantum gravity
and BH-thermodynamics.  In fact, this is the standard justification
for the extended use of the Wick-rotation.  On the other hand, when
the theory is formulated as a diffeomorphism invariant theory of
connections, one looses even the notion of a space-time metric; it
now becomes a {\it derived} notion and, therefore, the role of the
Wick rotation turns out to be not so obvious (see \cite{am2t}).
For, in the completely solvable models like $BF$ theories and $2+1$
gravity with no cosmological constant, the rigorously defined
measures on the space of histories have heuristic analogs involving
a factor of the form $\exp(iS)$ for both gauge groups ($SO(3)$ and
$SO(2,1)$), so the signature of the space-time metric seems to play
no role.  If one takes the viewpoint that the heuristic path
integrals based on the actions (\ref{actionL}, \ref{actionY}) fully
represent quantum gravity in three dimensions, then one is naturally
lead to conclude that the $H$-duality is, in a sense, more
fundamental and natural than the Wick transform.  For, the $H$-dual
theories not only share the space of histories, in their
Chern-Simons version, but also have the property that the actions
(\ref{actionL},\ref{actionY}), when evaluated on the shared history,
differ only by a sign \cite{aa:ac}.  Thus, {\it the path integral
for these two actions must be equivalent} (since the factor involved
is of the form $\exp(iS)$ and the actions are real).

\end{enumerate}

Implications of the previous discussion are intriguing. 
First, one can justify the
formal assignments of thermodynamical properties like entropy and
temperature to cases where, classically, there is no time-like KVF
and no horizon. This can be done by applying the standard
arguments of Gibbons-Hawking together
with the $H$-duality\cite{ac:ag2}.
Second, there
seems to be some tension between the relevance of the Wick transform
for quantum gravity in its metrical/geometrodynamical formulation 
and
the holonomy duality for its connection-dynamics formulation.
It would be of interest to fully understand the role of these
dualities in the context of path integrals for diffeomorphism
invariant theories. 
Finally, we could speculate about some implications for
the $H$-dual of the $AdS/CFT$ correspondence and what it could tell
us about quantum gravity in higher dimensions.

\section*{Acknowledgments}

The authors would like to thank the participants of the
15th Pacific Coast Gravity Meeting, February 1999, and
in particular  Abhay Ashtekar, M\'aximo Ba\~nados and Steven Carlip
for discussions. 
This research was supported in part by the National Science
Foundation under Grant No. PHY94-07194.
AC would like to thank the ITP for its
hospitality during which part of this work was completed. AC was 
also supported by DGAPA-UNAM Proy. No. IN121298 and by
CONACyT (M\'exico) Ref. No. I25655-E.
The work of AG was also supported in part by National Science
Foundation grant PHY97-22362, and by
funds from Syracuse University.

\begin{appendix}
\section*{Rotating Solutions: Boost with a Twist}
\label{appA}

\subsection{BTZ-$T^2$ Family}

Let us start by considering the Euclidean BTZ black hole with 
angular
momentum,
\begin{equation}
\d s^2=\left( -M+\frac{r^2}{l^2}-\frac{J^2}{4r^2}\right)\d \tau^2+
\left(-M+\frac{r^2}{l^2}-\frac{J^2}{4r^2}\right)^{-1}\d r^2+
r^2\left(\d\phi-\frac{J}{4r^2}\d \tau\right)^2
\label{Jnot0}
\end{equation}
where
\begin{equation}
r_\pm=\left\{\frac{Ml^2}{2}\left[
1\pm\left(1+\frac{J^2}{M^2l^2}\right)^{1/2}
\right]\right\}^{1/2}
\end{equation}
\begin{equation}
M=\frac{r_+^2+r_-^2}{l^2}\qquad ;\qquad J=\pm\frac{2r_+r_-}{l}
\end{equation} 
(recall that $8G=1$).
In order to construct this space-time, we can again start with the
 metric (\ref{met}),
\begin{equation}
\d s^2=l^2\left[\sinh^2\rho \,\d \chi^2+ 
\d \rho^2+\cosh^2\rho\,\d\varphi^2
\right]
\label{A3.18}
\end{equation}
where $\rho\in [0,\infty)$, $\chi\in[0,2\pi)$, 
$\varphi\in (-\infty,\infty)$. Note that 
$\xi=\frac{\partial}{\partial \varphi}$, and $\varphi$ is an 
affine
parameter of the finite ``boost" along the integral curves of
$\xi$.
The region that the embedding (\ref{btz:emb}) covers has the 
topology $S^1\times R^2$. 

The rotating BTZ solution is constructed by 
specifying the points that need to be identified in order
to have a two torus. Just as in the non-rotating case, one of
the generators is given by the (closed) orbits of the
$\partial/\partial\chi$ KVF, with period $2\pi$. Thus, one only
needs to specify the other generator. This is done by
periodically identifying the point
$(\varphi=0,\chi=0)$ with the point given by a boost along
$\xi$ with parameter $\frac{2\pi r_{+}}{l}$, 
together with a
rotation along the KVF $\frac{\partial}{\partial\chi}$
equal to $2\pi \frac{|r_-|}{l}$. Here $(r_+,|r_-|)$ are
the  parameters labeling the space-time. What one does is to
compactify along the line, in the plane $(\varphi,\chi)$, 
connecting,
say $(0,0)$ and $(\frac{2\pi r_{+}}{l},\frac{2\pi |r_-|}{l})$
and one ends up with a
space-time of topology $T^2\times R$.

We can define the new periodic coordinates 
$(\phi^\prime,\chi^\prime)$
along the generators of the torus,
\begin{equation}
\varphi=\T\frac{r_+}{l}\,\phi^\prime
\qquad;\qquad \chi=\chi^\prime+\T\frac{|r_-|}{l}\phi^\prime
\end{equation}
where $\chi^\prime\in  [0,2\pi)$ and $\phi^\prime\in [0,2\pi)$.
Then, the line element has the form,
\begin{equation}
\d s^2=l^2\left[\sinh^2\rho(\d \chi^\prime + 
\T\frac{|r_-|}{l}\d\phi^\prime)^2
+\d \rho^2+\cosh^2\rho(\T\frac{r_+}{l}\d\phi^\prime)^2\right]
\end{equation}
In order to put it in the standard form, let us define new 
coordinates
\begin{equation}
\chi^\prime=:\T\frac{(r_+^2+|r_-|^2)}{l^2r_+}\,\tau
\qquad
\phi^\prime=:\phi-\T\frac{|r_-|}{lr_+}\,\tau
\end{equation}
where $\phi\in[0,2\pi)$ and 
$\tau\in[0,\frac{2\pi l^2r_+}{r_+^2+|r_-|^2})$.
Then, the metric takes the ``proper radial
coordinates" form \cite{BBO},
\begin{equation}
\d s^2= \sinh^2\rho\left(\T\frac{r_+}{l}\d \tau +|r_-
|\d\phi\right)^2 + l^2\d\rho^2+
\cosh^2\rho\left(\T\frac{|r_-|}{l}\d\tau-r_+\d\phi\right)^2
\end{equation}
Finally, if we define $r^2:=r_+^2\cosh^2\rho+|r_-|^2\sinh^2\rho$,
we recover the standard form (\ref{Jnot0}).

In terms of
the gauge invariant information, i.e., the coordinates of 
$\hat{\Gamma}$ given by the four parameters $(a_1,s_1,a_2,s_2)$ 
\cite{ezawa,aa:ac},
a (Euclidean) BTZ black hole of mass $M$ and angular momentum
$J$ has coordinates in
$\hat{\Gamma}$ equal to $(2\pi,0,(2\pi |r_-|)/l,(2\pi r_+)/l)$.

The corresponding Lorentzian  dual space-times
are constructed using the same holonomy parameters. Starting from 
the metric (\ref{bang}),
\begin{equation}
\d s^2 = -l^2 \d t^2 + l^2\cosh^2 t \ \d\chi^2 + l^2 \sinh^2 t \
\d\varphi^2 \ .
\label{bang2}
\end{equation}
Along the $\chi$ direction we keep the $2\pi$ periodicity, and the
other generator is constructed via a boost in the 
$\partial/\partial\varphi$ direction with parameter $2\pi b$
together with a rotation along $\partial/\partial\chi$ with 
parameter
$2\pi a$. Thus, the spacetime has 
has coordinates in
$\hat{\Gamma}$ equal to $(2\pi,0,2\pi a,2\pi b)$. It is the
holonomy dual to the BTZ black hole when we set
$a=\frac{|r_-|}{l}$ and $b=\frac{r_+}{l}$.
We can define the new periodic coordinates 
$(\varphi^\prime,\chi^\prime)$
along the generators of the torus,
\begin{equation}
\varphi=b\,\varphi^\prime
\qquad;\qquad \chi=\chi^\prime+a\varphi^\prime
\end{equation}
where $\chi^\prime\in  [0,2\pi)$ and $\phi^\prime\in [0,2\pi)$.
Then, the line element has the form,
\begin{eqnarray}
\d s^2 &=& l^2\left[-\d t^2+
\cosh^2t(\d \chi^\prime + 
a\d\phi^\prime)^2
+\sinh^2t(b\d\phi^\prime)^2\right] \\
&=& -l^2 \d t^2 + l^2 \cosh^2 t | \d\chi + z(t) \d \phi|^2 \ ,
\end{eqnarray}
where $z(t) = a +ib\tanh t$ is the so called modulus of the torus, 
which 
completely determines the conformal geometry of it. For
$t\rightarrow\infty$, 
this parameter goes to the constant $a+ib$, and the torus will 
evolve
blowing up 
with the conformal factor $\exp(-2t)$. For $t=0$ the torus 
degenerates
to a circle of radius $2\pi(1+a)$.

Finally, the line element can be put in the form,
\begin{equation}
\d s^2=\frac{-\d T^2}{\left((b^2-a^2)+\frac{T^2}{l^2}-
\frac{a^2b^2}{T^2}\right)}+\left((b^2-a^2)+\T\frac{T^2}{l^2}-
\T\frac{a^2b^2}{T^2}\right)\d R^2+T^2(\d\phi+\T\frac{ab}{2}
\d R)^2
\end{equation}
that can be heuristically obtained from (\ref{Jnot0}) by
the substitution $l\mapsto il$, $\tau\mapsto it$,
$J\mapsto iJ$ and by the relabeling $t\mapsto R$ and
$r\mapsto T$.

The reduced phase space for a $T^2\times R$ topology space-time
is four dimensional and parameterized by $(s_i,a_i)$. However,
the BTZ black hole family spans only a two dimensional surface
(given by $r_+$ and $|r_-|$, with $0\leq |r_-|\leq r_+$ ),
while the Lorentzian cosmological solutions
are well behaved for any point on $\hat{\Gamma}$. What are then,
the Euclidean duals to the cosmological solutions that are off
the BTZ torus? These solutions have been studied by Carlip and
Teitelboim \cite{car:teit} who refer to them as 
``off-Shell" black holes. They are (Euclidean) $T^2\times R$
solutions, with a conical singularity and a twist at the
$\rho=0$ horizon. Thus, one can not include this point
in the (maximally evolved) space-time and obtain a regular
space-time with the $S^1\times R^2$ topology of the black holes. 
If one did not ask for regularity at the horizon and only asked
for a space-time defined in the region $\rho > 0$ (as in the
cosmological case for $T>0$), the ``off-shell" black holes would
be perfectly valid, as solutions to the Einstein equations.

\subsection{Rotating Particles}

Let us first construct the Euclidean space-times corresponding to
a `rotating particle' at the origin. These spaces are the Wick
transforms of the (naked singularity) Lorentzian space-times with
a rotating particle at the origin. The topology of these 
space-times,
both Lorentzian and Euclidean is $S^1\times R$. Thus, we only need 
to
specify the holonomy along the unique non-trivial loop. The explicit
procedure for the construction of the spacetime includes some
cutting and gluing together, starting from
 the Euclidean De-Sitter space. Recall that in
the case of zero angular momentum, the construction involved cutting
a wedge, removing the points inside the wedge, and gluing together 
the
lines that defined the original wedge (for a constant time).
Now, when angular momentum is
present, we still cut a wedge, but now we shall identify points that
are not only related by a rotation but also a boost. Note that
the particular rotation and boost that we have chosen in the
construction of
this Appendix are such that they leave invariant orthogonal 
planes, which implies that, as elements of
the group $SO(3,1)$, they commute. 
The starting point is again the region of $M^4$ parameterized by
(\ref{4.17}) and with line element,
\begin{equation}
\d s^2=l^2\left[\cosh^2\rho\, \d {t^{\prime}}^2+ 
\d \rho^2+\sinh^2\rho\,\d\phi^2
\right]\label{4.18bis}
\end{equation}
where $\rho\in [0,\infty)$, $\phi\in[0,2\pi)$, 
$t^{\prime}\in (-\infty,\infty)$.
Here, the KVF along which the identification will be made are
$\zeta:=\frac{\partial}{\partial\phi}$ and
$\xi:=\frac{\partial}{\partial t^{\prime}}$. We again cut a wedge
with angle $2\pi\alpha$ and identify points that differ by a 
rotation
by $2\pi\alpha$ along $\zeta$ and a boost with parameter $2\pi a$ 
along
$\xi$. In order to implement these identifications, let us define the
new coordinates
\begin{equation}
t:=t^{\prime}-\T\frac{a}{\alpha}\phi\quad;\quad\varphi:=
\T\frac{\phi}{\alpha}
\end{equation}
then,
\begin{equation}
\d s^2=l^2\left[\cosh^2\rho \,(\d t+a \d \varphi)^2+ 
\d \rho^2+\sinh^2\rho\,(\alpha\d\varphi)^2
\right]\label{4.18b}
\end{equation} 
where $t\in(-\infty,\infty)$, $\varphi\in[0,2\pi)$ and
$\rho\in(0,\infty)$. Now, we define the new coordinates
$r:=l\alpha\sinh\rho$ and $\tau:=(l/\alpha)t$, we get,
\begin{equation}
\d s^2=\left(\alpha^2+\frac{r^2}{l^2}\right)(\d\tau+
{\T\frac{la}{\alpha}\d\varphi)}+\left(\alpha^2+\frac{r^2}{l^2}\right
)^{-1
}
\d r^2+r^2\d\varphi^2
\end{equation}
note that this is the Wick transform of the AdS-point particle
solution given in \cite{brown}.

The construction of the Lorentzian solution follows the same steps.
The starting point are the embeddings (\ref{3.26}) and the induced
metric in De-Sitter,
\begin{equation}
\d s^2 = -l^2  \cos^2\chi \ \d\tau^2 + l^2\d\chi^2 +
l^2 \sin^2\chi \ \d\phi^2 \ . 
\label{part2bis}
\end{equation}
Again, we cut a wedge of size $2\pi\alpha$ and identify using a 
rotation
by $2\pi\alpha$ along $\zeta$ together with a boost along $\xi$ of
parameter $2\pi a$. With the new coordinates
$\tilde{\tau}:=\tau-(a/\alpha)\phi$ and $\varphi:=(l/\alpha)\phi$ we
get,
\begin{equation}
\d s^2=-l^2\cos^2\chi(\d\tilde{\tau}+a\d\varphi)^2+l^2\d\chi^2+
l^2\sin^2\chi(\alpha\d\varphi)^2
\end{equation}
which becomes, after defining $r:=l\alpha\sin\chi$,
$t:=(l/\alpha)\tilde{\tau}$,
\begin{equation}
\d s^2=-\left(\alpha^2-\frac{r^2}{l^2}\right)(\d 
t+{\T\frac{la}{\alpha}}
\d\varphi)^2+  \left(\alpha^2-\frac{r^2}{l^2}\right)^{-1}\d r^2+r^2
\d\varphi^2
\end{equation}

 \end{appendix}

\end{document}